\documentclass[twocolumn,showpacs,preprintnumbers,prl,amsmath,amssymb,superscriptaddress]{revtex4}
\usepackage{graphicx}
\usepackage{endnotes}
\usepackage{bm}
\usepackage{epsfig}
\usepackage{color}
\usepackage{upgreek}
\usepackage{ulem}

\preprint{Ver. 2.10}

\begin{document}


\title{Nuclear magnetic resonance at up to 10.1 Giga-Pascal pressure detects an electronic topological transition in aluminum metal}


\author{Thomas~Meissner}
\affiliation{Faculty of Physics and Earth Science, University of Leipzig, Linn\'{e}strasse 5, D-04103 Leipzig, Germany}
\author{Swee~K.~Goh}
\affiliation{Cavendish Laboratory, University of Cambridge, J.J. Thomson Avenue, Cambridge CB3 0HE, United Kingdom}
\author{J\"{u}rgen~Haase}
\affiliation{Faculty of Physics and Earth Science, University of Leipzig, Linn\'{e}strasse 5, D-04103 Leipzig, Germany}

\author{Manuel~Richter}
\author{Klaus~Koepernik}
\author{Helmut~Eschrig\footnote[2]{Deceased during writing of the paper.}}
\affiliation{IFW Dresden, P.O. Box 270116, D-01171 Dresden, Germany}

\date{\today}


\begin{abstract}
High-sensitivity $^{27}$Al nuclear magnetic resonance (NMR) measurements of aluminum metal under high pressures of up to 10.1 GPa reveal an unexpected negative curvature in the pressure-dependence of the electronic density of states measured through shift and relaxation, which violates free electron behavior. A careful analysis of the Fermiology of aluminum shows that pressure induces an electronic topological transition (Lifshitz transition) that is responsible for the measured change in the density of states. The experiments also reveal a sudden increase in the NMR linewidth above 4.2~GPa from quadrupole interaction, which is not in agreement with the metal's cubic symmetry.
\end{abstract}


\maketitle

The fundamental properties of metals are well known \cite{Ashcroftbook}, but nuclear magnetic resonance (NMR) at Giga-Pascal pressures supported by band structure calculations open a new window for their deeper understanding. This will be demonstrated here with a surprising result for an almost free electron system, simple aluminum metal -- the most abundant metal of the earth's crust.

Already in 1936, a decade before the discovery of NMR, Heitler and Teller \cite{Heitler1936} recognized that the relaxation rate $(1/T_1)$ at which nuclei exchange energy with the lattice is much faster in metals than in non-metals due to the high density of electronic states (DOS) at the Fermi level, $N(E_{\rm F})$. Later in 1949, Knight found a large NMR frequency shift (Knight shift) caused by electron spin paramagnetism \cite{Knight1949, Townes1950}, and Korringa \cite{Korringa1950} related $1/T_1$ to the temperature ($T$) independent Knight shift $K_S$. His famous relation, $T_1TK_S^2= \left(\gamma_e/\gamma_n\right)^2\hbar/(4\pi k_B)$, depends only on the gyromagnetic ratios of nucleus $(\gamma_n)$ and electron $(\gamma_e)$, apart from fundamental constants. Material properties may enter as electronic correlation effects through a correction factor \cite{Pines1954}. 

Qualitative theories frequently describe $N(E)$ with simple analytic functions. However, in real systems $N(E)$ is always non-analytic at those energies where the band structure, $E(\bf k)$, has extrema or saddle points. These details can profoundly influence $N(E_{\rm F})$, which is responsible for many characteristic properties of a metal, including conductivity, specific heat, or the critical temperature of superconductivity.

In an interesting scenario van Hove singularities in $N(E)$ can be shifted by varying parameters like chemical composition, magnetic field, or pressure. If a singularity passes through the Fermi level the metal undergoes an electronic topological transition (ETT), often referred to as Lifshitz transition, with a change in topology of the Fermi surface (FS)~\cite{lifshitz60}. Unfortunately, changes in the chemical composition can induce disorder that masks subtle effects~\cite{varlamov89}, and the required magnetic field strengths often cannot be reached with normal magnets~\cite{kozlova05}. Therefore, varying pressure appears to be the most appropriate tool for such studies (e.g., \cite{rosner06}). 

The compressibility of Al metal  \cite{Ashcroftbook} tells us that for a 10\% volume reduction, for which one might expect substantial effects, we need a pressure of about 7.6~GPa. Such pressures can routinely be achieved with anvil cells. However, experimental methods that investigate the electronic structure are rare under such conditions since in anvil cells two conical anvils with sub-millimeter culet size push on the tiny sample and the metallic gasket that encloses it. For example, in NMR experiments one needs to record a very weak, precessing nuclear magnetization with a resonant radio frequency (RF) coil that should fit the sample tightly, which was not possible with anvil cells, so far \cite{Conradi}.

A variety of different NMR setups have been employed in the past to investigate materials under such high pressure conditions \cite{Lee1987, Bertani1992, Pravica1998, Okuchi2005, Kitagawa2010}. Recently, a new high-sensitivity NMR anvil cell was introduced by some of us that uses an RF microcoil inside the high pressure region \cite{Haase2009}, similar to designs used for low-frequency measurement of the susceptibility \cite{Alireza03} and the de Haas-van Alphen effect \cite{Goh08} at high pressure. While this method has great potential in other areas (as was shown, e.g., with the closing of the NMR pseudogap of a high temperature superconductor \cite{Meissner2011}), its combination with bandstructure calculations seems to be particularly fruitful for the investigation of metals.

Piston-cylinder type Moissanite anvil cells  made of non-magnetic high-strength beryllium copper (BeCu) were loaded in Cambridge and readied for NMR in Leipzig. Metallic aluminum powder (Alfa Aesar, purity 99.97\%, particle size less than $44~\upmu {\rm m}$) was filled into a 9 to 10 turn RF microcoil wound from $12~\upmu{\rm m}$ diameter copper wire that was placed inside the hole of the BeCu-gasket. The microcoil also contained a small ruby chip for pressure monitoring \cite{Jayaraman1986}, Daphne 7373 and Glycerin were the pressure transmitting fluids. Various cells were used with home-built NMR probes that fit our standard wide-bore NMR magnets. Typical NMR experiments with $\pi/2$ pulse lengths of less than $1~\upmu{\rm s}$ were employed to excite and record free induction decays for shift and relaxation measurements. All measurements were carried out at room temperature.
\begin{figure}
  \centering
  \includegraphics[width=0.45\textwidth]{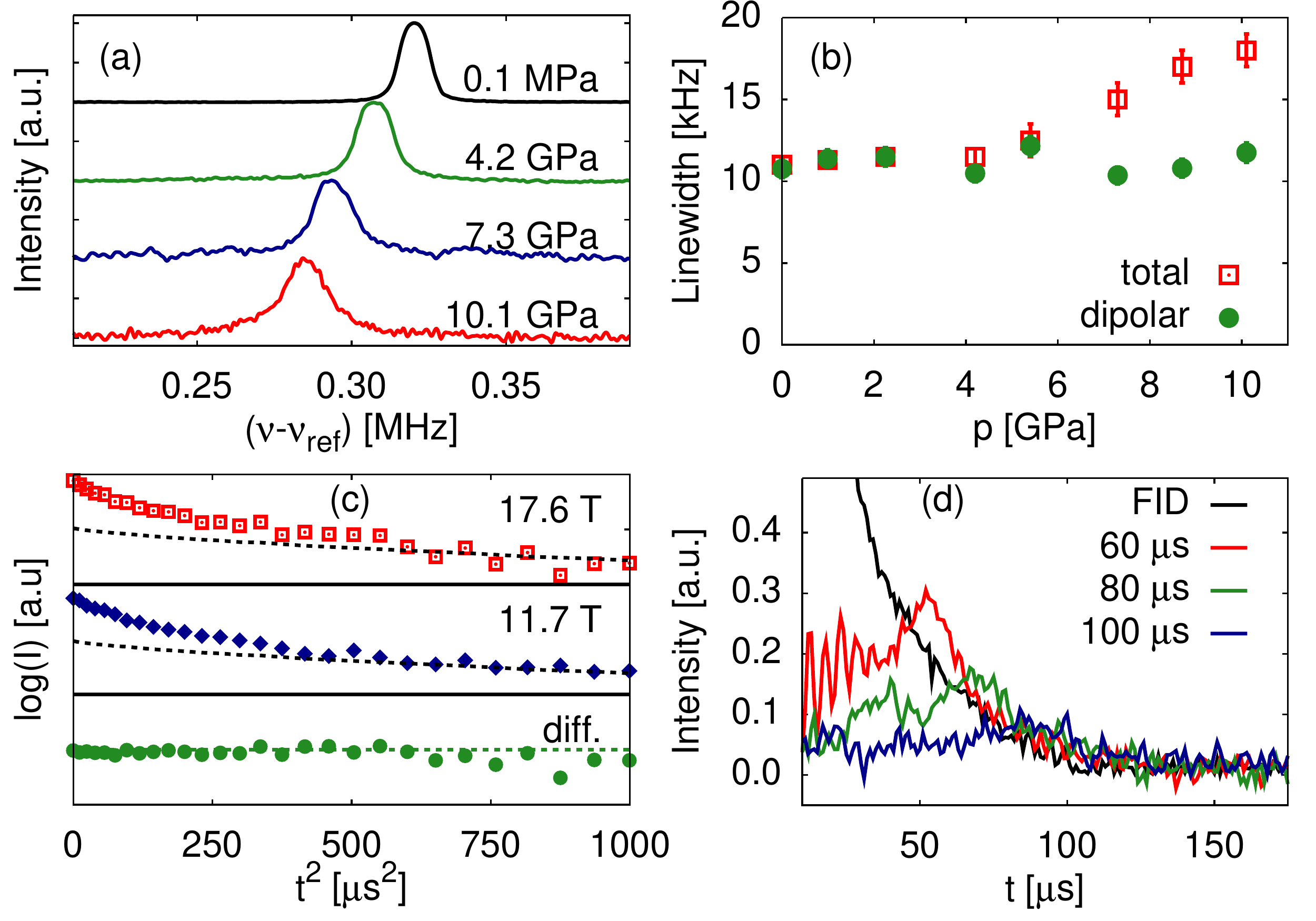}
  \caption{(Color online) $^{27}$Al NMR of aluminum metal. (a) Spectra at 17.6~T as a function of pressure, (b) Pressure dependence of the total linewidth (squares) and the dipolar contribution (circles). (c) Field dependence of the FID at 10.1~GPa (symbols) and ambient pressure (dashed line). The difference is defined as $\log[I(17.6~{\rm T})]\! - \! \log[I(11.7~{\rm T})]$. (d) FID and Hahn echoes of varying pulse separations ($60,\ 80,\ 100~\upmu{\rm s}$) observed at 5.4~GPa.}
  \label{fig1}
\end{figure}

Some typical {$^{27}$Al} NMR spectra up to 10.1~GPa together with other experimental results are shown in Fig.~\ref{fig1}. In panel (a) we see spectra as a function of pressure at 17.6~T. One observes a gradual shift of the resonance to lower frequency and an increase in the linewidth (and loss in intensity) as the pressure increases. The $^{27}$Al nuclei in the metal are in a cubic environment and one expects only a single resonance line for this $I=5/2$ quadrupolar nucleus. At ambient pressure, we find a Knight shift $K_{S}\! =\! (\nu_0\! -\! \nu_{\rm ref})/\nu_{\rm ref}$ of $1640~{\rm ppm}$ (the reference was AlCl$_3\cdot n$H$_2$O), in good agreement with what has been reported before \cite{Andrew1973}. We find the free induction decay (FID) at ambient pressure to be nearly Gaussian, cf.\ dashed lines in Fig.~\ref{fig1}~(c), with a second moment of 9.86~G$^2$, as expected for homonuclear dipolar coupling and in agreement with the literature value of 9.8~G$^2$ \cite{Spokas1959}.  (Beats in the FID at longer times, which were observed early on \cite{Spokas1959} and that are similar to those observed in CaF$_2$ \cite{Lowe1957} can be explained in terms of properties of chaotic many-body quantum systems \cite{Fine2004, Meier2012}).

In terms of the Korringa relation, we find at 295~K that $T_1T\! \approx\! 1.85~{\rm sK}$, also in agreement with earlier studies that showed that the relation holds with the same constant $1.85\pm0.05~{\rm sK}$ between about 1~K and 1000~K \cite{Spokas1959}. 

In Fig.~\ref{fig1} (b) we show the total linewidth (squares) and the dipolar contribution 2$\sigma\sqrt{2\log(2)}/\pi$ to it (circles), where $\sigma$  is the second moment determined from the beginning of the FID. As the pressure increases we observe that the dipolar linewidth does not change (from the change in lattice constants \cite{Syassen} we estimate a $10\%$ increase), while the total width shows a rather steep increase above about 4~GPa to almost twice its ambient pressure value at 10.1~GPa.

While no structural phase transition is expected to occur in our pressure range \cite{Akahama06}, we cannot exclude quadrupolar effects for {$^{27}$Al} NMR, e.g., from mechanical strain. In fact, in such a case an increase in linewidth is expected. Furthermore, a shift of the resonance frequency might occur due to the quadrupole interaction impairing the exact determination of the Knight shift. 

\begin{figure}
  \centering
  \includegraphics[width=0.45\textwidth]{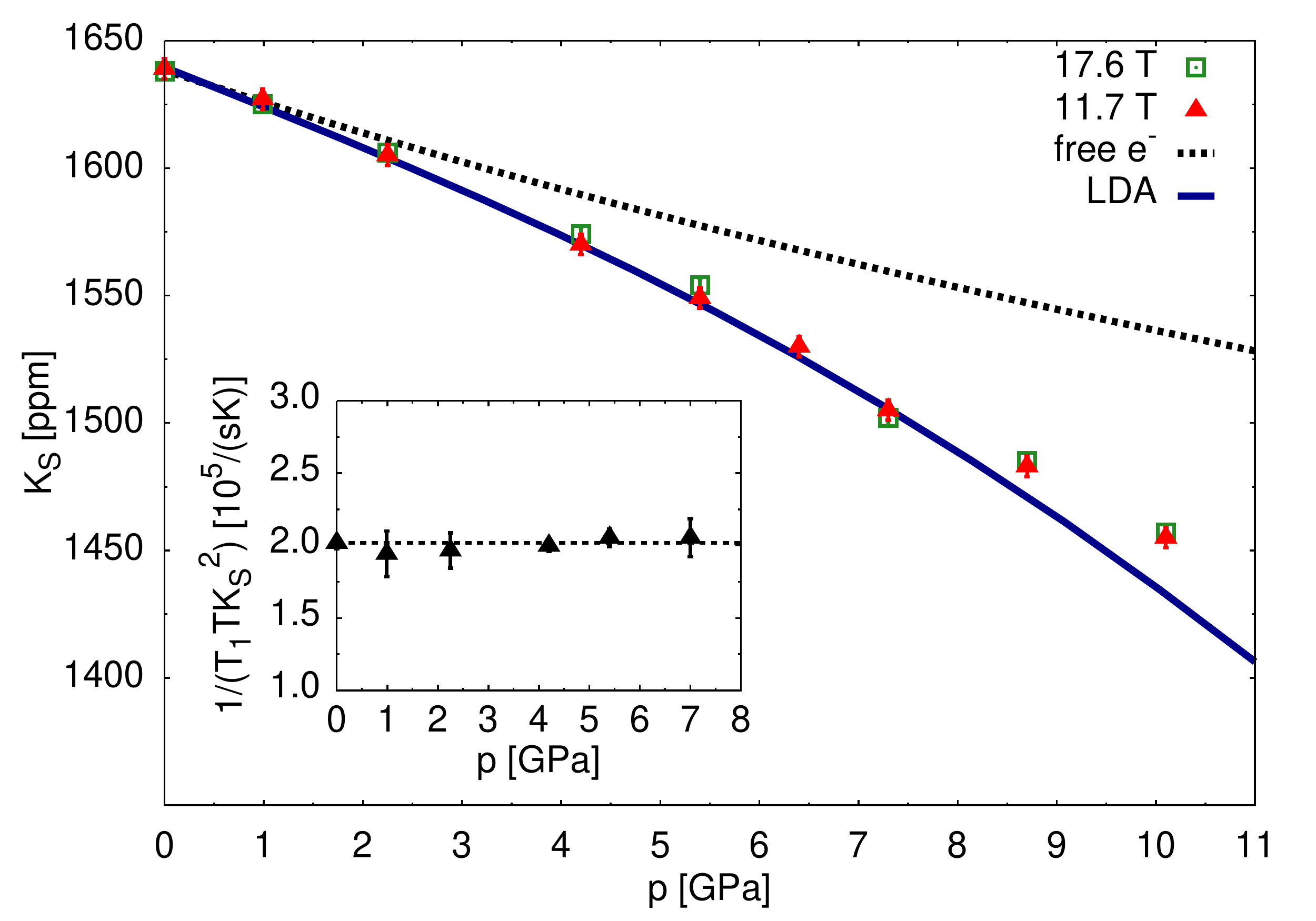}
  \caption{(Color online) Experimentally observed Knight shift $K_{S}$ at 11.7~T (triangles) and 17.6~T (squares). Also shown is the pressure dependence of $K_S$ for a free electron gas (dashed line) and that predicted from the LDA calculations. (Inset) Pressure dependence of the Korringa factor $1/(T_1TK_S^2)$.}
  \label{fig2}
\end{figure}

In order to prove that the observed pressure-dependent shift is of magnetic origin we performed NMR experiments at various magnetic field strengths ($B_0=$ 7.0~T, 11.7~T, and 17.6~T). Sample FIDs at 10.1~GPa are shown in {Fig.~\ref{fig1}~(c)}. We find that the shifts are indeed purely magnetic, cf.\ also Fig.~\ref{fig2}, and that both linewidths contributions are independent of  field. With the nuclear magnetic dipole broadening being independent of field we conclude that the additional broadening must be quadrupolar. We estimate the quadrupolar broadening to be about 13.5~kHz at 10.1~GPa, cf.\ {Fig.~\ref{fig1}~(b)}. In such a case one also expects that one can excite a weak Hahn spin echo that is absent for pure dipolar coupling \cite{Haase1993}. Indeed, we find no Hahn echo at ambient pressure, but the appearance of a weak echo at high pressures, cf.\ Fig.~\ref{fig1} (d). While we are not sure about the origin of this weak quadrupolar coupling at high pressures, now, we are certain that we can neglect it while interpreting the changes in shift and relaxation. 

The pressure dependence of the Knight shift, $K_{S}(p)$, (also for different magnetic field strengths) is shown in the main panel of Fig.~\ref{fig2}. Also shown (dotted line), is the expected dependence ($K_{S}(p) \! \propto \! N(E_{\rm F})\! \propto \! V^{2/3}$) for a free electron metal of volume $V$ \cite{Syassen}, which is clearly in contradiction with our measurements. Note that we do not expect the hyperfine coupling constant to change considerably with pressure as discussed below. The spin lattice relaxation at all pressures was found to be mono exponential, but pressure dependent. In the inset of Fig.~\ref{fig2} we plot the Korringa relation determined from our shift and relaxation data. It proofs that our observed changes in $K$ and $T_1$ can indeed be understood in terms of the hyperfine coupling to the s-electrons.

In order to shed light on the discovered discrepancy between free electron behavior of the shift and the experimental data, we decided to perform numerical bandstructure calculations. 
The calculations were carried out with the full-potential local-orbital scheme, FPLO-9.00-34~\cite{koepernik99} using the local density approximation (LDA Perdew-Wang 92) and a scalar-relativistic mode. The DOS was evaluated by linear-tetrahedron integrations with Bl\"ochl corrections on an extremely fine mesh of $300 \times 300 \times 300$ points in the full Brillouin zone (BZ) in order to achieve the required accuracy of the DOS close to singularities.

\begin{figure}\centering
\includegraphics[scale=0.36]{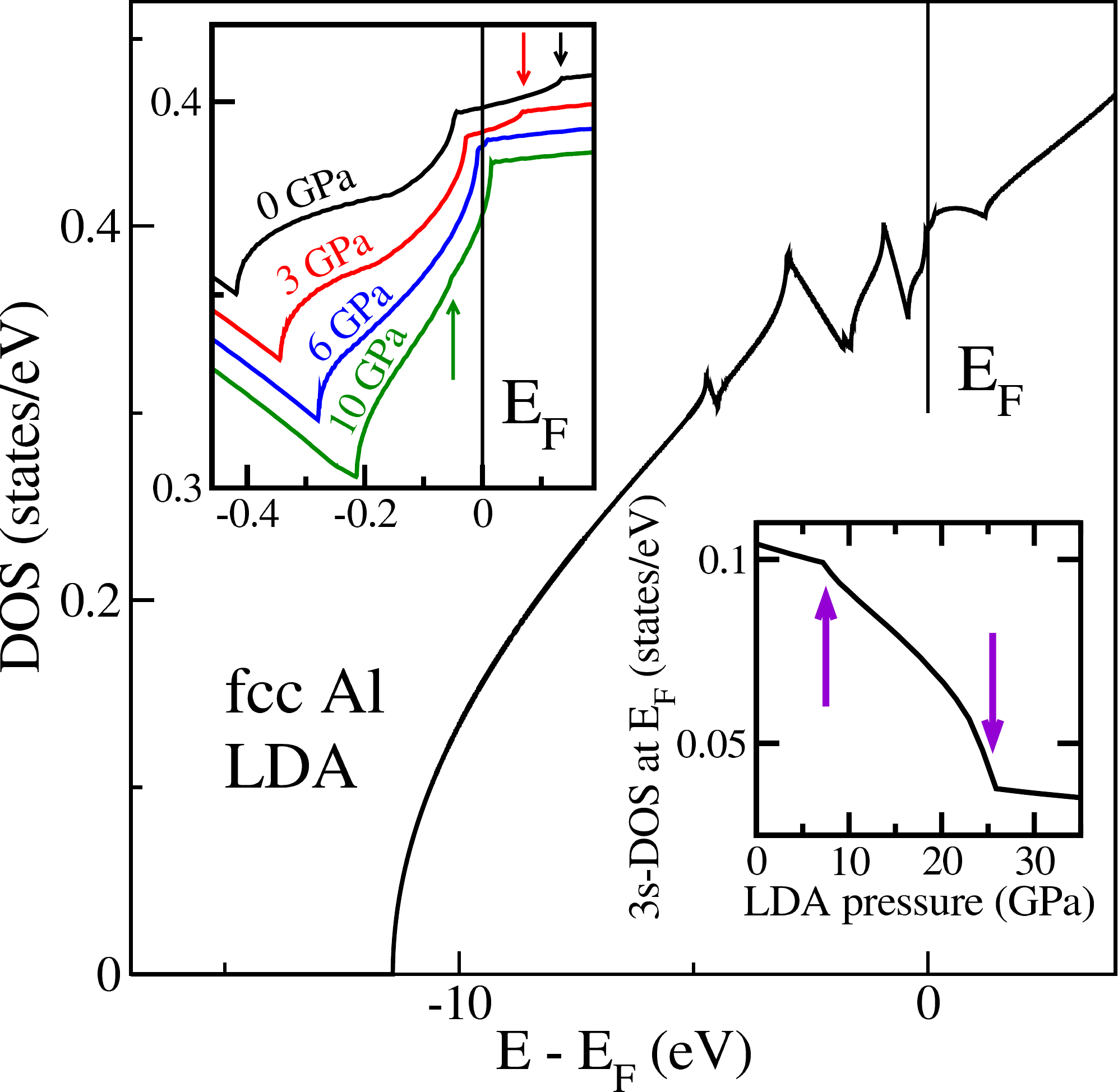}
\caption{\label{fig3} (Color online)
Electronic DOS of face centered cubic aluminum in LDA. (Main panel) Zero-pressure DOS, the Fermi energy $E_{\rm F}$  is indicated by a vertical line. (Upper inset) Zoom near $E_{\rm F}$ for different pressures (see text). (Lower inset) Pressure dependence of $N_{\rm 3s}(E_{\rm F})$, arrows indicate the positions of two ETT.}
\end{figure}

Fig.~\ref{fig3} shows in the main panel the calculated total electronic DOS. From the band bottom to about -5 eV we find almost perfect free-electron behavior, but a number of van-Hove singularities appear near the Fermi level. The pressure evolution of the total DOS close to $E_{\rm F}$ with higher energy resolution is depicted in the upper inset (the equation-of-state was obtained from a standard Birch-Murnaghan fit resulting in a zero-pressure LDA volume of 15.831 \AA{}$^3$, a bulk modulus of 83.7 GPa, and its pressure derivative 4.4). In the occupied part two singularities move upward in energy with increasing pressure, while a weak singularity in the unoccupied part moves downward (arrows). In the lower inset, the 3s DOS at $E_{\rm F}$ ($N_{\rm 3s}(E_{\rm F}))$ is shown as a function of pressure. Note that $N_{\rm 3s}(E_{\rm F})$ is non-analytic at those pressures where the topology of the FS changes and related van-Hove singularities pass $E_{\rm F}$. The two visible critical points at about 7.5 GPa and at about 25 GPa indicated by arrows are related to the two stronger singularities discussed above, while the third and weaker one is not resolved in this figure (it  almost merges with the singularity at 7.5 GPa). 

In order to compare the calculated DOS with the room-temperature NMR experiments, $N_{\rm 3s}(E_{\rm F})$ was folded with a 300 K Fermi-Dirac function and the zero pressure point was normalized to $K_S(0)$ (Fig.~\ref{fig2}). As one can see, the experimental and theoretical data match almost perfectly. In the FPLO code, distortion of the wave functions under pressure is taken into account by including atomic-like wave functions beyond a minimum basis set. In the present case, the valence basis comprised 2s, 2p, 3s, 3p, 3d, 4s, and 4p states. If a 4s contribution is taken into account, the high-pressure side of the theoretical curve shifts upward by about 6 ppm (in units of $K_S$). This means, we find an increase of the hyperfine coupling (averaged over 3s and 4s contributions) of 0.4\% at 10 GPa, compared to its zero-pressure value. Semicore 2s and core 1s contributions are found to be negligible.

Qualitatively, the critical point at 7.5 GPa readily explains the measured negative curvature of $K_S(p)$: pressure drives the system through an ETT and the pressure dependence of $N_{\rm 3s}(E_{\rm F})$ is stronger on the high-pressure side of the transition than on its low-pressure side.
\
\begin{figure}\centering
\includegraphics[scale=0.30]{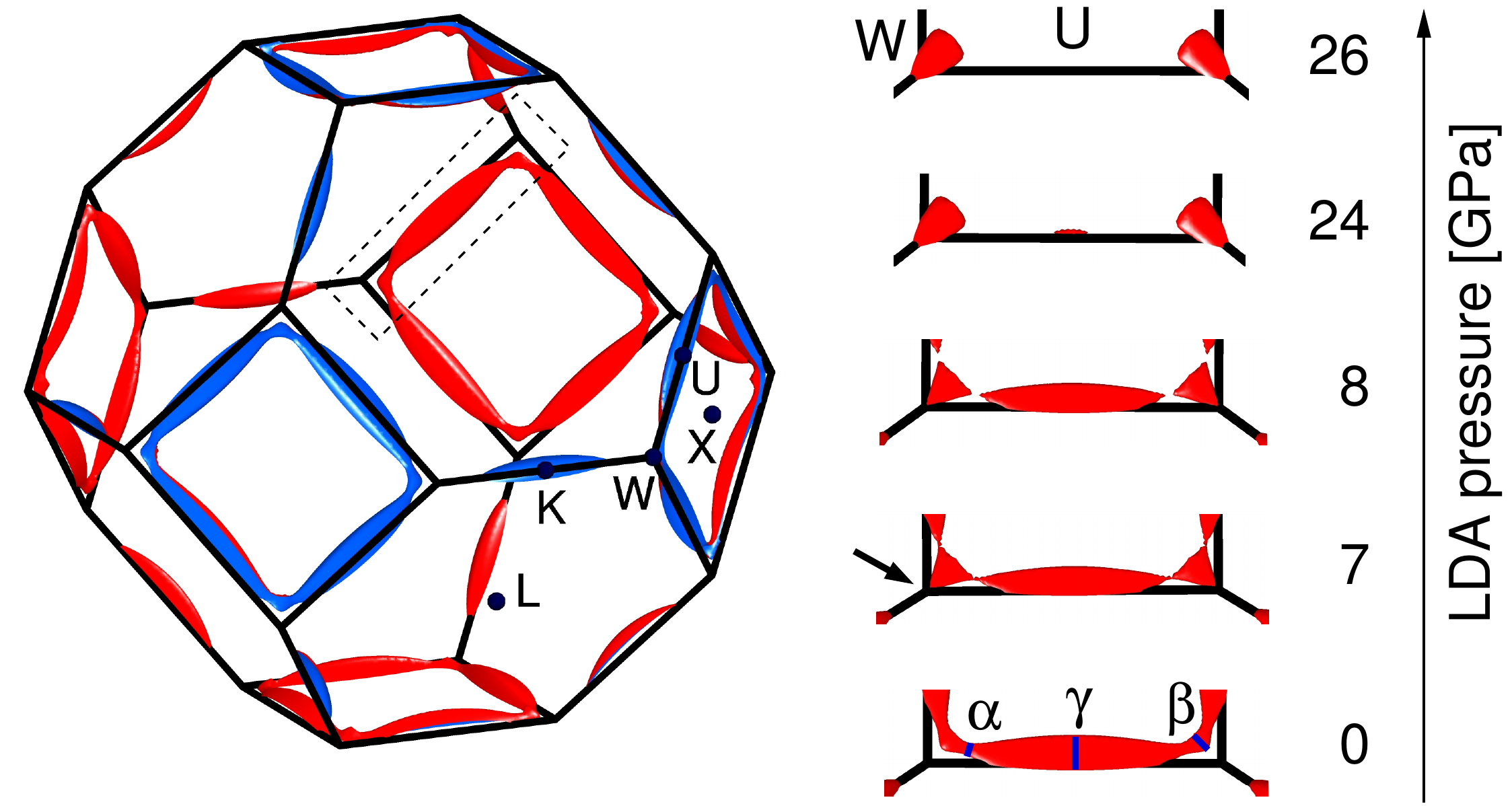}
\caption{\label{fig4} (Color online) Calculated Fermi surface of aluminum. Blue (red) color indicates the occupied (unoccupied) side. Left part: The well-known third-zone electron monster at zero pressure. Right part: Detail of the FS with pressure dependence (indicated at the right-hand side). At zero pressure (lowermost figure), the detail is identical with the framed part of the left figure. Extremal orbits in the common notation~\cite{joss80} are indicated.  The $\alpha$-orbit vanishes at an ETT between 7 and 8~GPa  and the $\gamma$-orbit vanishes between 24 and 26~GPa.}
\end{figure}

Fig.~\ref{fig4} illustrates the origin of the DOS singularities in the pressure dependence of the third-zone FS of Al. At $p=0$ the textbook four-armed ``monster'' is to be seen stretching along the square around X. It is characterized by three extremal orbits $\alpha$, $\beta$, and $\gamma$. At about 7 GPa, the monster has developed spikes which now touch the point W (arrow in Fig.~\ref{fig4}; a saddle point of the band structure at W crosses the Fermi level). Since another spike touches W from the neighboring BZ, the Fermi surface becomes multiply connected at this ETT~\cite{bross04}. While this topology change certainly has a drastic effect on magnetotransport, it creates only a minor singularity in the DOS (arrows in the upper left insert to Fig. \ref{fig3}) and, thus, is not visible in the Knight shift.

Only slightly above 7~GPa the monster breaks into parts. At this transition, the $\alpha$-orbit vanishes and $N_{\rm 3s}(E_{\rm F})$ shows a downward-kink (right lower insert to Fig. \ref{fig3}). From this kink, $K_S(p)$ obtains the observed negative curvature. Continued increase of the pressure lets the arms (or, spindles) around U shrink further until they vanish in another ETT. This happens at about 25 GPa and causes an upward-kink in $N_{\rm 3s}(E_{\rm F})$, which should manifest itself in a positive curvature of $K_S(p)$ at $p\approx 25$ GPa that could not be reached with NMR experiments yet.

Early studies of the FS of Al suggested that it is fairly well described by the free-electron model, see Ref. \cite{harrison59} and references therein. Later, de Haas-van Alphen experiments at pressures up to $0.7$ GPa hinted at deviations from that model~\cite{melz66}. No such deviations were found in Fermi momenta measured by positron annihilation up to 10 GPa~\cite{burton68}, but the error bars associated with the reported data are sizable. Existing experimental information was used by Joss and Monnier to fit pseudopotential parameters and, subsequently, to predict the general linear strain dependence of Al FS areas~\cite{joss80}. In accordance with this parameterization, an ETT under uniaxial stress close to 0.5 GPa has been identified by magneto-thermopower measurements on Al whiskers~\cite{overcash81}. Further, experimental data on Al alloyed with 6-10\% Si have been interpreted in terms of an ETT~\cite{livanov02}. We are not aware, however, of any previous experimental observation of ETT in Al under hydrostatic pressure. Our calculated transition pressures for the vanishing of the $\alpha$- and $\gamma$-orbits are by factors of two to four lower than what one would estimate from the zero-pressure derivatives of the FS areas obtained by Joss and Monnier~\cite{joss80}. Thus, linear approximations for the pressure dependence of the FS area are obviously not in general valid for pressures beyond 1 GPa.\\

To conclude, we have studied the Knight shift and relaxation of aluminum metal as a function of pressure up to 10.1 GPa, extending the pressure range previously investigated by a factor of 14 \cite{kushida_pressure_1971}. (Note that the authors in \cite{kushida_pressure_1971} found an increase in Knight-shift with pressure that we do not observe. However our first pressure point at 1 GPa is already higher in pressure than any of the data of the previous study.) Up to 10.1 GPa, we found a decrease of the Knight shift by 11\% which is almost three times of what is expected for a free electron system. Since we also observed an increase in linewidths with pressure (above 4 GPa) that is inconsistent with homonuclear dipolar interactions we performed an extensive study of the field dependence of the linewidths in order to exclude quadrupolar effects on the shift. We find the additional linewidth to be independent of field and we ascribe it to a small quadrupole interaction (13~kHz) that appears rather suddenly above 4~GPa. Since we used different anvil cells and different pressure transmitting fluids we believe that the measured, weak quadrupole coupling is not simply caused by strain, but a manifestation of a charge density variation that breaks the cubic symmetry, as suggested earlier \cite{Meissner2010}; further experiments are currently under way to study this effect.

Since the pressure dependent shift is magnetic and the Korringa relation does not change with pressure, we attribute the changes in $K_S(p)$ and $T_1(p)$ to changes in the electronic DOS. We performed numerical bandstructure calculations with high resolution and find quantitative agreement with the experimental data. We also find that the behavior is due to a kink in the electronic DOS at an electronic topological transition. While such transitions are normally only observable at low temperatures, our NMR data taken on a large enough pressure interval allow the detection of ETT even at room temperature. We predict that a further increase in pressure (up to 30 GPa) will result in a further drop of the Knight shift by more than a factor of two. Finally, we expect a positive curvature of $^{27}K(p)$ around $p=25$ GPa due to the disappearance of the spindle-like Fermi surface sheet close to the U-point.


\begin{acknowledgements}
The authors acknowledge financial support by the University of Leipzig, the IRTG: Diffusion in porous media, Royal Society and Trinity College (Cambridge), as well as discussions with Marvin Cohen (Berkeley), Bernd Rosenow (Leipzig), Mick Mantle (Cambridge), Chris Pickard (London), Stefan-Ludwig Drechsler, and Michael D. Kuz'min (Dresden).
\end{acknowledgements}

\end{document}